\documentclass[12pt]{iopart}

\usepackage{amssymb}
 \usepackage{amsthm}

 \newtheorem{theorem}{Theorem}[section]

\newcommand{\ud}{\mathrm{d}}

\begin{document}

\title[]{Nonextensive Pythagoras' Theorem}



\author{Ambedkar Dukkipati}

\address{EURANDOM, P.O. Box 513, 5600 MB Eindhoven, The Netherlands}
\ead{\mailto{dukkipati@eurandom.tue.nl}}

\begin{abstract}
        Kullback-Leibler relative-entropy, in cases involving distributions
        resulting from relative-entropy minimization, has a
        celebrated property reminiscent of squared Euclidean distance:
        it satisfies an analogue of the Pythagoras' theorem. And
        hence, this property is referred to as Pythagoras' theorem of
        relative-entropy minimization or triangle equality and 
        plays a fundamental role in geometrical approaches of
        statistical estimation theory like information geometry.
        Equvalent of Pythagoras' theorem in the generalized
        nonextensive formalism is established in {\bf (Dukkipati at
        el., Physica A, 361 (2006)
        124-138~\cite{DukkipatiMurtyBhatnagar:2006:NonextensiveTriangleEquality})}. In
        this paper we give a detailed account of it.
\end{abstract}

\maketitle

\section{Introduction}
\label{Section:Introduction}
        \noindent
        Apart from being a fundamental measure of information, 
        Kullback-Leibler relative-entropy or KL-entropy plays a role
        of `measure of the distance' between two probability
        distributions in statistics. Since it is not a metric, at first
        glance, it might seem that the geometrical interpretations
        that metric distance measures provide usually might not be possible
        at all with the KL-entropy playing a role as a distance measure on
        a space of probability distributions. But
        it is a pleasant surprise that it is possible to formulate
        certain geometric propositions for probability distributions, with
        the relative-entropy playing 
        the role of squared Euclidean distance. Some of these
        geometrical interpretations cannot be derived from the 
        properties of KL-entropy alone, but from the properties of ``KL-entropy
        minimization''; restating the previous statement, these
        geometrical formulations are possible only when probability
        distributions resulting from ME-prescriptions of
        KL-entropy are involved.

        As demonstrated by Kullback~\cite{Kullback:1959:InformationTheoryAndStatistics}, 
        minimization problems of relative-entropy
        with respect to a set of moment constraints find their
        importance in the well known {\em Kullback's minimum
        entropy principle} and thereby 
        play a basic role in the information-theoretic approach to
        statistics~\cite{Good:1963:MaximumEntropyForHypothesisFormulation,IrelandKullback:1968:ContingencyTablesWithGivenMarginals}. 
        They frequently occur elsewhere also, e.g., in the theory 
        of large
        deviations~\cite{Sanov:1957:OnTheProbabilityOfLargeDeviationsOfRandomVariables},
        and in statistical physics, as maximization of
        entropy~\cite{Jaynes:1957:InformationTheoryStatisticalMechanics_I,Jaynes:1957:InformationTheoryStatisticalMechanics_II}.

        Kullback's minimum entropy principle can be considered as a
        general method of inference about an 
        unknown probability distribution when there exists a prior
        estimate of the distribution and new information in the form of
        constraints on expected
        values~\cite{Shore:1981:PropertiesOfCrossEntropyMinimization}.
        Formally, one can state this principle as:
        given a prior distribution $r$, of all the 
        probability distributions that satisfy the given moment
        constraints, one should choose the
        \textit{posterior} $p$ with the least relative-entropy.
        The prior distribution $r$ can be a 
        {\it reference} distribution (uniform, Gaussian, Lorentzian
        or Boltzmann etc.) or a {\it prior} estimate of $p$.
        The principle of Jaynes maximum entropy is a special case of
        minimization of relative-entropy under appropriate
        conditions~\cite{ShoreJohnson:1980:AxiomaticDerivationOfThePrincipleMaxEntMinEnt}.

        Many properties of relative-entropy minimization just reflect
        well-known properties of relative-entropy but there are
        surprising differences as
        well. For example, 
        relative-entropy does not generally satisfy a triangle relation
        involving three arbitrary probability distributions. But in
        certain important cases involving distributions that result
        from relative-entropy minimization, relative-entropy results
        in a theorem comparable to the Pythagoras' theorem
        cf.~\cite{Csiszar:1975:I-devergenceOfProbabilityDistributionsAndMinimizationProblems}
        and
        \cite[\S~11]{Cencov:1982:StatisticalDecisionRulesAndOptimalInference}. 
        In this geometrical interpretation, relative-entropy plays the
        role of squared distance 
        and minimization of relative-entropy appears as the analogue
        of projection on a sub-space in a Euclidean geometry. This
        property is also known as triangle
        equality~\cite{Shore:1981:PropertiesOfCrossEntropyMinimization}. 

        The main aim of this paper is to study the possible
        generalization of Pythagoras' theorem to the nonextensive
        case. Before we take up this problem, we present the properties
        of Tsallis relative-entropy  minimization and present some
        differences with the classical case. In the representation of
        such a minimum entropy distribution, we highlight the
        use of the {\em $q$-product} ($q$-deformed version of
        multiplication), an operator that has 
        been introduced recently to derive the mathematical structure
        behind the Tsallis statistics.
        Especially, $q$-product representation of Tsallis minimum
        relative-entropy distribution will be useful for the
        derivation of the equivalent of triangle equality for Tsallis
        relative-entropy. We mention here that a general class of
        relative-entropy functionals which satisfy Pythagorean
        relation is established by Gr{\"{u}}nwald and
        Dawid~\cite{GrunwaldDawid:2004:GameTheoryMaximumEntropyMinimumDiscrepancy}.
        Recently a 
        Pythagoras' theorem for a version of
        R{\'{e}}nyi relative 
        entropy is reported by
        Sundaresan~\cite{Sundaresan:2006:GuessingUnderSourceUncertainty}.  

        Before we conclude this introduction on geometrical ideas
        of relative-entropy minimization, we make a note on the other
        geometric approaches.
        One approach is that 
        of Rao~\cite{Rao:1945:InformationAndAccuracyAttainable}, where one
        looks at the set of probability distributions on a
        sample space as a differential manifold and introduce a
        Riemannian geometry on this manifold. This approach is
        pioneered by
        {\v{C}}encov~\cite{Cencov:1982:StatisticalDecisionRulesAndOptimalInference} 
        and
        Amari~\cite{Amari:1985:DifferentialGeometricMethodsInStatistics}
        who 
        have shown the existence of a particular Riemannian geometry which is
        useful in understanding some questions of statistical
        inference. This Riemannian geometry turns out to have
        some interesting connections with information theory and as shown
        by Campbell~\cite{Campbell:1985:TheRelationBetweenInformationTheoryAndTheDifferentialGeometry},
        with the minimum relative-entropy. In this approach too, the
        above mentioned Pythagoras' Theorem plays an important
        role~\cite[pp.72]{AmariNagaoka:2000:MethodsOfInformationGeometry}.

        The other idea involves the use of Hausdorff dimension~\cite{Billingsley:1960:HausdorffDimensionInProbabilityTheory,Billingsley:1965:ErgodicTheoryAndInformation}
        to understand why minimizing relative-entropy should provide
        useful results. This approach was begun by
        Eggleston~\cite{Eggleston:1952:SetsOfFractionalDimention} for
        a special 
        case of maximum entropy and was developed by
        Campbell~\cite{Campbell:1992:MinimumRelativeEntropyAndHausdorffDimension}. 
        For an excellent review on various geometrical aspects associated with minimum
        relative-entropy one can refer
        to~\cite{Campbell:2003:GeometricIdeasInMinimumCross-Entropy}.



        The structure of this paper is organized as follows.
        We present the necessary background in
        \S~\ref{Section:PT:RelativeEntropyMinimization}, where we
        discuss properties of relative-entropy minimization in the
        classical case. In
        \S~\ref{Section:PT:TsallisRelativeEntropyMinimization}, we
        present the ME prescriptions of Tsallis relative-entropy and
        discuss its differences with the classical case. Finally, the
        derivation of Pythagoras' theorem in the nonextensive case is
        presented in
        \S~\ref{Section:PT:NonextensivePythagorasTheorem}.

        Regarding the notation, we define all the information measures 
        on the measurable space $(X,\mathfrak{M})$. The default reference
        measure is $\mu$ unless otherwise stated. 
        For simplicity in exposition, we will not
        distinguish between functions differing on a $\mu$-null set
        only; nevertheless, we can work with equations between
        $\mathfrak{M}$-measurable functions on $X$ if they are
        stated as being valid only $\mu$-almost everywhere ($\mu$-a.e or
        a.e).
        Further we assume that all the quantities of interest
        exist and also assume, implicitly, the $\sigma$-finiteness of $\mu$ and
        $\mu$-continuity of probability measures whenever
        required. Since these assumptions repeatedly occur in various
        definitions and formulations, these will not be mentioned in
        the sequel.
        With these assumptions we do not distinguish between 
        an information measure of pdf $p$ and that of the corresponding probability
        measure $P$ -- hence when we give definitions of
        information measures for pdfs, we also use  the corresponding
        definitions of probability measures as well, wherever 
        convenient or required  --  with the understanding that $P(E) = \int_{E} p\,
        \ud \mu $, and the converse holding as a result of the
        Radon-Nikodym theorem, with $p = 
        \frac{\ud P}{\ud \mu}$. In both the cases we have $P \ll \mu$.
        
        Note that though results presented in this
        paper do not 
        involve major measure theoretic concepts, we write all the
        integrals with respect to the measure $\mu$, as a convention; these integrals
        can be replaced by summations in the discrete case or Lebesgue 
        integrals in the continuous case.

\section{Relative-Entropy Minimization in the Classical Case}
\label{Section:PT:RelativeEntropyMinimization}
        \noindent
        Kullback's minimum entropy principle can stated
        formally as follows. Given a prior distribution $r$ with a
        finite set of moment constraints of the form 
        \begin{equation}
        \label{Equation:PT:ExpectationConstraints}
        \int_{X} u_{m}(x) p(x) \, \ud \mu(x) =
        \langle u_{m} \rangle \enspace, \:\:\:m = 
        1, \ldots , M \enspace,
        \end{equation}
        one should choose the posterior $p$ which minimizes
        the relative-entropy
        \begin{equation}
         \label{Equation:PT:RelativeEntropy_of_pdf} 
        I(p\|r) = \int_{X} p(x) \ln \frac{p(x)}{r(x)} \, \ud \mu(x) \enspace.
        \end{equation}
        In (\ref{Equation:PT:ExpectationConstraints}), $\langle
        u_{m} \rangle$, $m=1,\ldots, M$ are the 
        known expectation values of  $\mathfrak{M}$-measurable functions $u_{m}: X
        \rightarrow \mathbb{R}, \:\: m = 1, \ldots, M$ respectively.

        With reference to (\ref{Equation:PT:RelativeEntropy_of_pdf})
        we clarify here that, though we mainly use expressions of
        relative-entropy defined 
        for pdfs in
        this paper,
        we use expressions in terms of corresponding 
        probability measures as well. For example, when we write
        the Lagrangian for relative-entropy minimization below, we use the
        definition of relative-entropy
         \begin{equation}
        I(P\|R) = \left\{ \begin{array}{ll}
        \displaystyle{\int_{X} \ln \frac{\ud P}{\ud R} \, \ud P }     &
        \:\:\:\:\:\textrm{if}\:\:\:\:\:  P \ll R \enspace, \\ \\
          +\infty   & \:\:\:\:\:\textrm{otherwise.}
           \end{array} \right.
        \end{equation}
        for
        probability measures $P$ 
        and $R$,
        corresponding to pdfs $p$ and $r$ respectively. This
        correspondence between probability measures $P$ and $R$ with pdfs
        $p$ and $r$, respectively, will not be described again in the sequel.
  \subsection{Canonical Minimum Entropy Distribution}
  \label{SubSection:PT:CanonicalMinimumEntropyDistribution}
        \noindent
        To minimize the
        relative-entropy~(\ref{Equation:PT:RelativeEntropy_of_pdf})
        with respect to the
        constraints~(\ref{Equation:PT:ExpectationConstraints}), the
        Lagrangian turns out to be
        {\setlength\arraycolsep{0pt}
        \begin{eqnarray}
        \label{Equation:PT:LagranginForRelativeEntropy}
        \mathcal{L}(x, \lambda, \beta) = \int_{X} \ln
        \frac{\ud P}{\ud R}(x)&&  \, \ud P(x)  + \lambda \left(\int_{X}\, \ud P(x) - 1
        \right) \nonumber \\
        && + \sum_{m=1}^{M} \beta_{m} \left(\int_{X} u_{m}(x)\, \ud P(x) -
        \langle u_{m} \rangle \right) \enspace,
        \end{eqnarray}}
        where $\lambda$ and $\beta_{m}, \:\: m=1, \ldots M$ are
        Lagrange multipliers.
        The solution is given by
        \begin{displaymath}
         \ln \frac{\ud P}{\ud R}(x) + \lambda + \sum_{m=1}^{M}
        \beta_{m} u_{m}(x) = 0 \enspace,
        \end{displaymath}
        and the solution can be written in the form of
        \begin{equation}
        \label{Equation:PT:RelativeEntropyDistribution_InTermsOfMeasures}
        \frac{\ud P}{\ud R} (x) = \frac{\displaystyle e^{ - \sum_{m=1}^{M} \beta_{m} 
        u_{m}(x)}}{\displaystyle \int_{X} e^{ - \sum_{m=1}^{M} \beta_{m} 
        u_{m}(x)} \, \ud R} \enspace.
        \end{equation}
        Finally, from
        (\ref{Equation:PT:RelativeEntropyDistribution_InTermsOfMeasures})
        the 
        posterior distribution $p(x) = \frac{\ud P}{\ud \mu}$ given by
        Kullback's minimum entropy 
        principle can be written in terms of the prior $r(x) = \frac{\ud
        R}{\ud \mu}$ as
        \begin{equation}
        \label{Equation:PT:MinimumCrossEntropyDistribution}  
          p(x) = \frac{r(x) e^{-\sum_{m=1}^{M} \beta_{m} u_{m}(x)} }
        {\widehat{Z}} \enspace, 
        \end{equation}
        where
        \begin{equation}
        \widehat{Z} = \int_{X} r(x) e^{-\sum_{m=1}^{M}
        \beta_{m} u_{m}(x)  } \, \ud \mu(x) 
        \end{equation}
        is the partition function. 

        Relative-entropy minimization has been applied to many
        problems in
        statistics~\cite{Kullback:1959:InformationTheoryAndStatistics}
        and statistical
        mechanics~\cite{Hobson:1971:ConceptsInStatisticalMechanics_SecondaryRef}.
        The other applications include pattern
        recognition~\cite{ShoreGray:1982:MinimumCrossEntropyPatternClassification},
        spectral 
        analysis~\cite{Shore:1981:MinimumCrossEntropySpectralAnalysis_SecondaryRef},
        speech
        coding~\cite{MarkelGray:1976:LinearPredictionOfSpeach_SecondaryRef},
        estimation of prior distribution for Bayesian
        inference~\cite{CatichaPreuss:2004:MaximumEntropyAndBayesianDataAnalysis}
        etc. 
        For a list of references on applications of relative-entropy
        minimization
        see~\cite{ShoreJohnson:1980:AxiomaticDerivationOfThePrincipleMaxEntMinEnt}
        and a recent paper
        \cite{Cherny:2004:OnMinimizationAndMaximizationOfEntropy}.


        Properties of relative-entropy minimization have been studied
        extensively and
        presented by Shore~\cite{Shore:1981:PropertiesOfCrossEntropyMinimization}.
        Here we briefly mention a few. 

        The principle of maximum entropy is equivalent to
        relative-entropy minimization in the special case of discrete
        spaces and uniform priors, in the sense that, when the prior is
        a uniform distribution with finite support $W$ (over $E
        \subset X$), the minimum entropy
        distribution turns out to be  
        \begin{equation}
        \label{Equation:PT:MinimumCrossEntropyDistribution_WithUniformPrior}
        p(x) = \frac{\displaystyle e^{ -\sum_{m=1}^{M} \beta_{m}
        u_{m}(x)}}{\displaystyle \int_{E} e^{
        -\sum_{m=1}^{M} \beta_{m} u_{m}(x)} \, \ud \mu(x)}  \enspace, 
        \end{equation}
        which is in fact, a maximum entropy distribution
        of Shannon
        entropy with respect to the 
        constraints (\ref{Equation:PT:ExpectationConstraints}).

        The important relations to relative-entropy
        minimization are as follows. 
        Minimum relative-entropy, $I$, can be calculated as
        \begin{equation}
        \label{Equation:PT:MinimumRelativeEntropy}
        I = - \ln \widehat{Z} - \sum_{m=1}^{M}
        \beta_{m} {\langle{{u}_{m}}\rangle}  \enspace,
        \end{equation}
        while the thermodynamic equations are
        \begin{equation}
        \label{Equation:PT:KL_ThermodyamicEquation_1}
        \frac{\partial}{\partial \beta_{m}} \ln \widehat{Z}  = -
        {\langle{{u}_{m}}\rangle} \enspace,\:\:\: m = 1, \ldots M,
        \end{equation}
        and
        \begin{equation}
        \label{Equation:PT:KL_ThermodyamicEquation_3} 
        \frac{\partial I}{\partial {\langle{{u}_{m}}\rangle}}
        =  - \beta_{m} \enspace, 
        \:\:\: m =1, \ldots M.
        \end{equation}
  \subsection{Pythagoras' Theorem}
        \noindent
        The statement of Pythagoras' theorem of relative-entropy
        minimization can be formulated as 
        follows~\cite{Csiszar:1975:I-devergenceOfProbabilityDistributionsAndMinimizationProblems}.
        \begin{theorem}
         \label{Theorm:PT:PTinClassicalCase_1} 
        Let $r$ be the prior, $p$ be the probability
        distribution that minimizes the relative-entropy subject to
        a set of constraints
        \begin{equation}
        \label{Equation:PT:InTheorem_PTinClassicalCase_1_ExpectationConstraints}
        \int_{X} u_{m}(x) p(x) \, \ud \mu(x) = \langle u_{m} \rangle \enspace,\, m = 
        1, \ldots , M \enspace,
        \end{equation}
        with respect to $\mathfrak{M}$-measurable functions $u_{m}: X
        \rightarrow \mathbb{R}$, $m = 1, \ldots M$ whose expectation
        values $\langle u_{m} \rangle, \, m=1,\ldots M$ are (assumed
        to be) a priori known. Let $l$ be any
        other distribution 
        satisfying the same constraints
        (\ref{Equation:PT:InTheorem_PTinClassicalCase_1_ExpectationConstraints}), then we have the
        triangle equality
        \begin{equation}
        \label{Equation:PT:TriangleEquality}
        I(l\|r) = I(l\|p) + I(p\|r) \enspace.
        \end{equation}
        \end{theorem}
        \proof
        We have
        {\setlength\arraycolsep{0pt}
        \begin{eqnarray}
        \label{Equation:PT:InTheorem_PTinClassicalCase_1_Int_1}
        I(l\|r) &\:=\:& \int_{X} l(x) \ln \frac{l(x)}{r(x)} \, \ud \mu(x)
        \nonumber \\
        &\:=\:& \int_{X} l(x) \ln \frac{l(x)}{p(x)} \, \ud \mu(x)
        + \int_{X} l(x) \ln \frac{p(x)}{r(x)} \, \ud \mu(x) \nonumber \\
        &\:=\:& I(l \| p) + \int_{X} l(x) \ln \frac{p(x)}{r(x)} \, \ud \mu(x)
        \end{eqnarray}}
        From the minimum entropy distribution
        (\ref{Equation:PT:MinimumCrossEntropyDistribution}) we have
        \begin{equation}
        \label{Equation:PT:InTheorem_PTinClassicalCase_1_Int_2}
        \ln \frac{p(x)}{r(x)} = - \sum_{m=1}^{M} \beta_{m} u_{m}(x) -
        \ln \widehat{Z} \enspace.
        \end{equation}
        By substituting
        (\ref{Equation:PT:InTheorem_PTinClassicalCase_1_Int_2}) in
        (\ref{Equation:PT:InTheorem_PTinClassicalCase_1_Int_1}) we get
        {\setlength\arraycolsep{0pt}
        \begin{eqnarray}
        I(l\|r) &\:=\:& I(l\|p) + \int_{X} l(x) \left\{  - \sum_{m=1}^{M} \beta_{m} u_{m}(x) -
        \ln \widehat{Z}  \right\} \, \ud \mu(x) \nonumber \\
        &\:=\:& I(l\|p) - \sum_{m=1}^{M} \beta_{m} \left\{ \int_{X} l(x) u_{m}(x)
        \, \ud \mu(x) \right\} - \ln \widehat{Z} \nonumber \\
        &\:=\:& I(l\|p) - \sum_{m=1}^{M} \beta_{m} {\langle u_{m} \rangle}
        - \ln \widehat{Z} \qquad \qquad \mbox{(By hypothesis)}
        \nonumber \\
        &\:=\:& I(l\|p)+ I(p\|r) \enspace. \qquad \qquad \mbox{(By
        (\ref{Equation:PT:MinimumRelativeEntropy}))} \nonumber
        \end{eqnarray}}
        \endproof

        A simple consequence of the above theorem is that
        \begin{equation}
        \label{Equation:PT:ConsequenceOfPythogoreanTheorem}
        I(l\|r) \geq I(p\|r)
        \end{equation}
        since $I(l\|p) \geq 0$ for every pair of pdfs, with equality if and only if
        $l =p$.

        Detailed discussions on the importance of Pythagoras' theorem of
        relative-entropy minimization can be found
        in~\cite{Shore:1981:PropertiesOfCrossEntropyMinimization} and
        \cite[pp.
        72]{AmariNagaoka:2000:MethodsOfInformationGeometry}. For a
        study of relative-entropy minimization without the use of
        Lagrange multiplier technique and corresponding
        geometrical aspects, one can refer
  to~\cite{Csiszar:1975:I-devergenceOfProbabilityDistributionsAndMinimizationProblems}.

        Pythagorean realtion of relative-entropy minimization not only
        plays a fundamental role in geometrical approaches of statistical
        estimation
        theory~\cite{Cencov:1982:StatisticalDecisionRulesAndOptimalInference}
        and information
        geometry~\cite{Amari:1985:DifferentialGeometricMethodsInStatistics,Amari:2001:HierarchyOfProbabilityDistributions}
        but is also important
        for applications in which 
        relative-entropy minimization is used for purposes of pattern
        classification and cluster
        analysis~\cite{ShoreGray:1982:MinimumCrossEntropyPatternClassification}.

\section{Tsallis Relative-Entropy Minimization}
\label{Section:PT:TsallisRelativeEntropyMinimization}
        \noindent
        Unlike the generalized entropy measures, ME of generalized
        relative-entropies is not much addressed in the literature.
        Here, one has to mention the work 
        in~\cite{BorlandPlastinoTsallis:1998:InformationGainWithinNonextensiveThermostatistics},
        where the minimum relative-entropy distribution
        of Tsallis relative-entropy with respect to the constraints in
        terms of $q$-expectation values is given.

        In this section, we study several aspects of Tsallis
        relative-entropy minimization. First we derive the minimum
        entropy distribution in the case of $q$-expectation values
        (see (\ref{Equation:PT:q-ExpectationConstraints})) and 
        then in the case of normalized $q$-expectation values
        (see (\ref{Equation:PT:Normalized-q-ExpectationConstraints})). We 
        propose an elegant representation of these distributions by
        using $q$-deformed binary operator called $q$-product
        $\otimes_{q}$. This operator is defined in
        \cite{Borges:2004:ApossibleDeformedAlgebra} along similar
        lines as $q$-addition $\oplus_{q}$
        and $q$-subtraction $\ominus_{q}$.
        Since $q$-product
        plays an important role in nonextensive formalism, we
        include a detailed discussion on the $q$-product in this
        section. Finally, we study properties of 
        Tsallis relative-entropy minimization and its differences with
        the classical case. 
  \subsection{Generalized Minimum Relative-Entropy Distribution}
        \noindent
        To minimize Tsallis relative-entropy
        \begin{equation}
        \label{Equation:PT:TsallisRelativeEntropyOf-pdf}            
        I_{q}(p\|r) = - \int_{X} p(x) \ln_{q} \frac{r(x)}{p(x)}\, \ud
        \mu(x) 
        \end{equation}
        with respect to the set of constraints specified in terms of
        $q$-expectation values
         \begin{equation}
        \label{Equation:PT:q-ExpectationConstraints}
        \int_{X} u_{m}(x) {p(x)}^{q} \, \ud \mu(x) =
        {\langle u_{m} \rangle}_{q} \enspace, m = 
        1, \ldots , M, 
        \end{equation}
        the concomitant variational principle is given as follows: Define
        {\setlength\arraycolsep{0pt}
        \begin{eqnarray}
        \label{Equation:PT:LagranginForTsallisRelativeEntropy_q-Expectation}
        \mathcal{L}(x, \lambda, \beta) = \int_{X} \ln_{q}&&
        \frac{r(x)}{p(x)} \, \ud P(x) - \lambda \left(\int_{X}\, \ud P(x) - 1
        \right) \nonumber \\
        && - \sum_{m=1}^{M} \beta_{m} \left(\int_{X} {p(x)}^{q-1} u_{m}(x) \, \ud P(x) -
        {\langle u_{m} \rangle}_{q} \right) 
        \end{eqnarray}}
        where $\lambda$ and $\beta_{m}, \:\: m=1, \ldots M$ are
        Lagrange multipliers. Now set
        \begin{equation}
          \frac{\ud  \mathcal{L}}{\ud P} =
          0 \enspace.
        \end{equation}  
        The solution is given by
        \begin{displaymath}
        \ln_{q} \frac{r(x)}{p(x)} - \lambda - {p(x)}^{q-1}
        \sum_{m=1}^{M} \beta_{m} u_{m}(x) = 0\enspace,
        \end{displaymath}
        which can be rearranged by using the definition of
        $q$-logarithm $\ln_{q} x = \frac{x^{1-q}-1}{1-q}$ as
        \begin{displaymath}
        p(x) = \frac{\left[ {r(x)}^{1-q} - (1-q) \sum_{m=1}^{M} \beta_{m}
        u_{m}(x) \right]^{\frac{1}{1-q}}}{\left(\lambda(1-q) +
        1\right)^{\frac{1}{1-q}}} \enspace.    
        \end{displaymath}
        Specifying the Lagrange parameter $\lambda$ via the
        normalization $\int_{X} p(x) \, \ud \mu(x) =1$, one can 
        write Tsallis minimum relative-entropy distribution
        as~\cite{BorlandPlastinoTsallis:1998:InformationGainWithinNonextensiveThermostatistics}
        \begin{equation}
         \label{Equation:PT:GeneralizedMinimumCrossEntropyDistribution_1}  
          p(x) = \frac{\displaystyle {\left[{r(x)}^{1-q} - (1-q)
         \sum_{m=1}^{M} \beta_{m} u_{m}(x)
         \right]}^{\frac{1}{1-q}}}{\displaystyle 
        \widehat{Z_{q}}} \enspace,
         \end{equation}
        where the partition function is given by
        \begin{equation}
        \widehat{Z_{q}} = \int_{X} {\left[{r(x)}^{1-q} - (1-q)
         \sum_{m=1}^{M} \beta_{m} u_{m}(x)  \right]}^{\frac{1}{1-q}}
        \, \ud \mu(x) \enspace. 
        \end{equation}
        The values of the Lagrange parameters $\beta_{m}$, $m = 1, \ldots ,
        M$ are determined using the constraints 
        (\ref{Equation:PT:q-ExpectationConstraints}).


  \subsection{$q$-Product Representation for Tsallis Minimum Entropy Distribution}
        \noindent
        Note that the generalized relative-entropy
        distribution~(\ref{Equation:PT:GeneralizedMinimumCrossEntropyDistribution_1})
        is not of the form
        of its classical counterpart~(\ref{Equation:PT:MinimumCrossEntropyDistribution}) even if
        we replace the exponential with the $q$-exponential. But one
        can express~(\ref{Equation:PT:GeneralizedMinimumCrossEntropyDistribution_1})
        in a form similar to the classical case by
        invoking $q$-deformed binary operation called $q$-product.

        In the framework of $q$-deformed functions and operators
        a new multiplication, called
        $q$-product defined as
        \begin{equation}
        \label{Equation:PT:q-product}
        x \otimes_{q} y \equiv \left\{ \begin{array}{ll}
        \left( x^{1-q} + y^{1-q} -1 \right)^{\frac{1}{1-q}} & \qquad
        \textrm{if $x,y >0$,} \\
            & \qquad \textrm{$x^{1-q} + y^{1-q} -1 > 0$}\\
          0 & \qquad \textrm{otherwise.}
           \end{array} \right.
        \end{equation}
        This is first introduced 
         in~\cite{NivanenMehauteWang:2003:GeneralizedAlgebraWithinNonextensiveStatistics}
         and explicitly defined
         in~\cite{Borges:2004:ApossibleDeformedAlgebra} for satisfying
        the following equations:
        \begin{eqnarray}
         \ln_{q} (x \otimes_{q} y) & = & \ln_{q} x + \ln_{q} y
        \label{Equation:PT:q-logorithmProduct} \enspace,\\
         e_{q}^{x} \otimes_{q} e_{q}^{y} & = & e_{q}^{x+y} \label{Equation:PT:q-exponentialProduct} \enspace.
        \end{eqnarray}
        The $q$-product recovers the usual product in the limit $q
        \rightarrow 1$ i.e., 
         $\lim_{q \to 1} (x \otimes_{q} y) = xy $. The fundamental
         properties of the $q$-product $\otimes_{q}$ are almost the
         same as the usual product, and the distributive law does not hold
        in general, i.e., 
        \begin{displaymath}
        a (x \otimes_{q} y ) \neq ax \otimes_{q} y \:\:\: (a,x,y \in \mathbb{R})\enspace.
        \end{displaymath}
        Further properties of the $q$-product can be found in
         \cite{NivanenMehauteWang:2003:GeneralizedAlgebraWithinNonextensiveStatistics,Borges:2004:ApossibleDeformedAlgebra}.

        One can check the mathematical validity of the $q$-product by
        recalling the expression of the exponential 
        function $e^{x}$ 
        \begin{equation}
        \label{Equation:PT:ExponentialFunction}
        e^{x} = \lim_{n \rightarrow \infty} {\left(1 +
        \frac{x}{n}\right)}^{n} \enspace.
        \end{equation}
        Replacing the power on the right side
        of~(\ref{Equation:PT:ExponentialFunction}) by $n$ times 
        the $q$-product $\otimes_{q}$:
        \begin{equation}
        \label{Equation:PT:n-times-q-product}
        x^{{\otimes_{q}}^{n}} =  \underbrace{x \otimes_{q} \ldots
        \otimes_{q} x}_{n \: \: \mathrm{times}} \enspace,
        \end{equation}
        one can verify that~\cite{Suyari:2004:q-StirlingsFormulaInTsallisStatistics}
        \begin{equation}
        \label{Equation:PT:q-ExponentialFunction}
        e_{q}^{x} = \lim_{n \rightarrow \infty} {\left(1 +
        \frac{x}{n}\right)}^{{\otimes_{q}}^{n}} \enspace.
        \end{equation}
        Further mathematical significance of $q$-product is demonstrated
        in~\cite{SuyariTsukada:2005:LawOfErrorInTsallisStatistics}
        by discovering the mathematical structure of statistics based
        on the Tsallis formalism: law of error, $q$-Stirling's formula,
        $q$-multinomial coefficient and experimental evidence of
        $q$-central limit theorem.

        Now, one can verify the non-trivial fact that Tsallis minimum
        entropy distribution 
        (\ref{Equation:PT:GeneralizedMinimumCrossEntropyDistribution_1})
        can be expressed as~\cite{DukkipatiMurtyBhatnagar:2005:PropertiesOfKullback-LeiblerCrossEntropyMinimization}, 
         \begin{equation}
         \label{Equation:PT:GeneralizedMinimumCrossEntropyDistribution_2}  
         p(x) = \frac{\displaystyle r(x) \otimes_{q} e_{q}^{-
         \sum_{m=1}^{M} \beta_{m} u_{m}(x)  }}{\displaystyle
         \widehat{Z_{q}}}  \enspace, 
         \end{equation}
        where
        \begin{equation}
        \widehat{Z_{q}} = \int_{X} r(x) \otimes_{q}
         e_{q}^{ - \sum_{m=1}^{M} \beta_{m} u_{m}(x) } \, \ud \mu(x) .
        \end{equation}
        Later in this paper we see that this representation is
        useful in establishing properties of Tsallis relative-entropy
        minimization and corresponding thermodynamic equations. 

        It is important to note that the distribution in
        (\ref{Equation:PT:GeneralizedMinimumCrossEntropyDistribution_1})
        could be a (local/global) 
        minimum only if $q > 0$ and the Tsallis cut-off condition
        specified by Tsallis maximum entropy distribution is 
        extended to the relative-entropy case i.e., $p(x) = 0$ whenever 
        $\left[{r(x)}^{1-q} - (1-q)
        \sum_{m=1}^{M} \beta_{m} u_{m}(x) \right] < 0$.
       The latter
       condition is also required for the $q$-product representation of the
       generalized minimum entropy distribution.

         In this case, one can calculate minimum relative-entropy
         $I_{q}$ as
         \begin{equation}
        \label{Equation:PT:MinimumTsallisRelativeEntropy_wrt_q-Expectation}
        I_{q} = - \ln_{q} \widehat{Z_{q}} - \sum_{m=1}^{M}
        \beta_{m} {\langle{{u}_{m}}\rangle}_{q}  \enspace.
        \end{equation}

        To demonstrate the usefulness of $q$-product representation
         of generalized minimum entropy distribution we
         present the verification
         (\ref{Equation:PT:MinimumTsallisRelativeEntropy_wrt_q-Expectation}). 
         By using the property of $q$-multiplication
         (\ref{Equation:PT:q-exponentialProduct}), Tsallis minimum
         relative-entropy distribution
         (\ref{Equation:PT:GeneralizedMinimumCrossEntropyDistribution_2})  
         can be written as
         \begin{displaymath}
           p(x) \widehat{Z_{q}} = e_{q}^{- \sum_{m=1}^{M}
             \beta_{m}u_{m}(x) + \ln_{q}r(x)} \enspace.
           \end{displaymath}
         By taking $q$-logarithm on both sides, we get
         \begin{displaymath}
           \ln_{q} p(x) + \ln_{q} \widehat{Z_{q}} + (1-q) \ln_{q} p(x)
           \ln_{q} \widehat{Z_{q}} = - \sum_{m=1}^{M}
           \beta_{m}u_{m}(x) + \ln_{q} r(x)
         \end{displaymath}  
         By the property of
         $q$-logarithm $\ln_{q}\left(\frac{\displaystyle
         x}{\displaystyle y} \right)
              = y^{q-1}( \ln_{q}x - \ln_{q}y)$, we have
         \begin{eqnarray}
         \label{Equation:PT:IntEqIntheDerivationOfMinRelativeEntropy_q-Expectation}
          \ln_{q} \frac{r(x)}{p(x)} =  p(x)^{q-1} \left\{  \ln_{q} \widehat{Z_{q}} +  (1-q)
           \ln_{q} p(x) \ln_{q}\widehat{Z_{q}} +   \sum_{m=1}^{M}
           \beta_{m}u_{m}(x)  \right\} \enspace. \nonumber \\
         \end{eqnarray}
         By substituting
         (\ref{Equation:PT:IntEqIntheDerivationOfMinRelativeEntropy_q-Expectation}) in
         Tsallis relative-entropy
         (\ref{Equation:PT:TsallisRelativeEntropyOf-pdf}) we get
         \begin{displaymath}
          I_{q} = - \int_{X} p(x)^{q}  \left\{  \ln_{q} \widehat{Z_{q}} +  (1-q)
           \ln_{q} p(x) \ln_{q}\widehat{Z_{q}} +   \sum_{m=1}^{M}
           \beta_{m}u_{m}(x)  \right\} \, \ud \mu(x) \enspace.
         \end{displaymath}
         By~(\ref{Equation:PT:q-ExpectationConstraints}) and 
         expanding $\ln_{q} p(x)$ one can write $I_{q}$ in its final
         form as in
         (\ref{Equation:PT:MinimumTsallisRelativeEntropy_wrt_q-Expectation}).

        It is easy to verify the following thermodynamic equations for the
        minimum Tsallis relative-entropy:
        \begin{equation}
        \label{Equation:PT:GeneralizedKL_ThermodyamicEquation_1}
        \frac{\partial}{\partial \beta_{m}} \ln_{q} \widehat{Z_{q}}  = -
        {\langle{{u}_{m}}\rangle}_{q} \enspace,\:\:\: m = 1, \ldots M,
        \end{equation}
        \begin{equation}
        \label{Equation:PT:GeneralizedKL_ThermodyamicEquation_3} 
        \frac{\partial I_{q}}{\partial {\langle{{u}_{m}}\rangle}_{q}
        } =  - \beta_{m} \enspace, 
        \:\:\: m =1, \ldots M,
        \end{equation}
        which generalize thermodynamic equations in the classical
        case.

  \subsection{The Case of Normalized $q$-Expectations}
        \noindent
        In this section we discuss Tsallis relative-entropy
        minimization with respect to the constraints in the form of
        normalized $q$-expectations 
        \begin{equation}
        \label{Equation:PT:Normalized-q-ExpectationConstraints}  
        \frac{\int_{X} u_{m}(x) p(x)^{q} \, \ud \mu(x)}{\int_{X}
          p(x)^{q}\, \ud \mu(x)} = {\langle\langle u_{m} \rangle\rangle}_{q} \enspace, m = 
        1, \ldots , M.
        \end{equation}

        The variational principle for Tsallis
        relative-entropy minimization in this case is as below. Let
        {\setlength\arraycolsep{0pt}    
        \begin{eqnarray}
        \label{Equation:PT:LagranginForTsallisRelativeEntropy_Normalizedq-Expectation}
        \mathcal{L}(x, \lambda, \beta) = && \int_{X} \ln_{q}
        \frac{r(x)}{p(x)} \, \ud P(x) - \lambda \left(\int_{X}\, \ud P(x) - 1
          \right) \nonumber \\
          && - \sum_{m=1}^{M} \beta^{(q)}_{m} \left(\int_{X} {p(x)}^{q-1}
          \left(u_{m}(x) - {\langle\langle u_{m}  \rangle\rangle}_{q}
          \right) \, \ud P(x) \right) \enspace,
        \end{eqnarray}}
         where the parameters $\beta_{m}^{(q)}$ can be defined in terms
        of the true Lagrange parameters $\beta_{m}$ as
        \begin{equation}
           \beta_{m}^{(q)} = \frac{\displaystyle
           \beta_{m}}{\displaystyle \int_{X}
           p(x)^{q}\, \ud \mu(x)}   \enspace, \qquad m = 1, \ldots, M.
        \end{equation}
        This gives minimum entropy distribution as
        \begin{equation}
        \label{Equation:PT:Tsallis_minimumRelativeEntropyDistribution_For_Normalizedq-Expectations} 
        p(x) =  \frac{1}{\widehat{\overline{{Z}_{q}}}} \left[{r(x)}^{1-q} - (1-q)
        \frac{\sum_{m=1}^{M} \beta_{m} \left( u_{m}(x) -
        {\langle\langle {u}_{m} \rangle\rangle}_{q} \right)}{\int_{X}
        {p(x)}^{q}\, \ud \mu(x)}
        \right]^{\frac{1}{1-q}}
        \end{equation}
        where
        \begin{displaymath}
        \widehat{\overline{{Z}_{q}}} = \int_{X} {\left[{r(x)}^{1-q} - (1-q)
        \frac{\sum_{m=1}^{M} \beta_{m} \left( u_{m}(x) -
        {\langle\langle {u}_{m} \rangle\rangle}_{q} \right)}{\int_{X}
        {p(x)}^{q}\, \ud \mu(x)}
        \right]}^{\frac{1}{1-q}} \, \ud \mu(x) \enspace. 
        \end{displaymath}
        Now, the minimum entropy distribution
        (\ref{Equation:PT:Tsallis_minimumRelativeEntropyDistribution_For_Normalizedq-Expectations})
        can be expressed using the $q$-product (\ref{Equation:PT:q-product}) as
        \begin{equation}
         \label{Equation:PT:Tsallis_minimumRelativeEntropyDistribution_For_Normalizedq-Expectations_q-Expo}  
        p(x) = \frac{1}{\widehat{\overline{{Z}_{q}}}} \left\{  r(x)
        \otimes_{q} \exp_{q}\left(  \frac{\sum_{m=1}^{M} \beta_{m} \left( u_{m}(x) -
        {\langle\langle {u}_{m} \rangle\rangle}_{q} \right)}{\int_{X}
        {p(x)}^{q}\, \ud \mu(x)}   \right)  \right\} \enspace.
        \end{equation}

        Minimum Tsallis relative-entropy $I_{q}$ in this case satisfies
        \begin{equation}
        \label{Equation:PT:MinimumTsallisRelativeEntropy_Normalized-q-Expect}
        I_{q} = - \ln_{q}\widehat{\overline{{Z}_{q}}} \enspace,
        \end{equation}
        while one can derive the following thermodynamic equations: 
        \begin{equation}
        \frac{\partial}{\partial \beta_{m}} \ln_{q} \widehat{Z_{q}}  = -
        {\langle\langle{{u}_{m}}\rangle\rangle}_{q} \enspace, \:\:\: m = 1, \ldots M,
        \end{equation}
        \begin{equation}
        \frac{\partial I_{q}}{\partial
        {\langle\langle{{u}_{m}}\rangle\rangle}_{q}  }  =  -
        \beta_{m} \enspace, \:\:\: m =1, \ldots M,
        \end{equation}
        where
        \begin{equation}
        \ln_{q} \widehat{Z_{q}} = \ln_{q} \widehat{\overline{{Z}_{q}}}
        - \sum_{m=1}^{M} \beta_{m}
        {\langle\langle{{u}_{m}}\rangle\rangle}_{q} \enspace.
        \end{equation}

\section{Nonextensive Pythagoras' Theorem}
\label{Section:PT:NonextensivePythagorasTheorem}
        \noindent
        With the above study of Tsallis relative-entropy minimization,  
        in this section, we present our main result, Pythagoras'
        theorem or triangle equality
        (Theorem~\ref{Theorm:PT:PTinClassicalCase_1}) generalized to
        the nonextensive case. To present this result, we shall  discuss
        the significance of triangle 
        equality in the classical case. We
        restate Theorem~\ref{Theorm:PT:PTinClassicalCase_1}
        which is essential for the derivation of the triangle equality in the
        nonextensive framework.
  \subsection{Pythagoras' Theorem Restated}
  \label{SubSection:PT:PTrestated}
        \noindent
        Significance of the triangle equality is evident in the following
        scenario. Let $r$ be the prior estimate of the unknown
        probability distribution $l$, about which, the information in the
        form of constraints
        \begin{equation}
        \label{Equation:PT:InSS_PTrestated_1}
        \int_{X} u_{m}(x) l(x) \, \ud \mu(x)  =
        {\langle{u}_{m}\rangle}\enspace, \:\:\: m =1, \ldots M 
        \end{equation}
        is available with respect to the fixed functions $u_{m}$, $m = 1, \ldots,
        M$. The problem 
        is to choose a posterior estimate $p$
        that is in some sense the best estimate of $l$ given by the
        available information i.e., prior $r$ and the information in the
        form of expected values
        (\ref{Equation:PT:InSS_PTrestated_1}). Kullback's minimum
        entropy principle provides a general  
        solution to this inference problem and provides us the
        estimate~(\ref{Equation:PT:MinimumCrossEntropyDistribution})
        when 
        we minimize relative-entropy $I(p\|r)$ with respect to
        the constraints
        \begin{equation}
        \label{Equation:PT:InSS_PTrestated_2}
        \int_{X} u_{m}(x) p(x) \, \ud \mu(x)  =
        {\langle{u}_{m}\rangle}\enspace, \:\:\: m =1, \ldots M \enspace. 
        \end{equation}

        This estimate of posterior $p$ by Kullback's minimum entropy
        principle also offers the relation (Theorem~\ref{Theorm:PT:PTinClassicalCase_1})
        \begin{equation}
        \label{Equation:PT:InSS_PTrestated_3}
        I(l\|r) = I(l\|p) + I(p\|r) \enspace,
        \end{equation}
        from which one can draw the following conclusions. By
        (\ref{Equation:PT:ConsequenceOfPythogoreanTheorem}), 
        the minimum relative-entropy posterior estimate of 
        $l$ is not only logically consistent, but also closer to $l$, in
        the relative-entropy sense, that is the prior $r$. Moreover,
        the difference $I(l\|r) - I(l\|p)$ is exactly the
        relative-entropy $I(p\|r)$ between the posterior and the
        prior. Hence, $I(p\|r)$ can be interpreted as the amount of
        information provided by the constraints that is not inherent
        in $r$.

        Additional justification to use minimum
        relative-entropy estimate of $p$ with respect to the 
        constraints (\ref{Equation:PT:InSS_PTrestated_2})
        is provided by the
        following {\em expected value
        matching}
        property~\cite{Shore:1981:PropertiesOfCrossEntropyMinimization}.
        To explain this concept we restate our above estimation
        problem as follows. 

        For fixed functions $u_{m}, \: m = 1, \ldots M$, let the
        actual unknown distribution $l$ satisfy
        \begin{equation}
         \label{Equation:PT:InSS_PTrestated_4} 
         \int_{X} u_{m}(x) l(x)\, \ud \mu(x)  =  {\langle{w}_{m}\rangle}
        \enspace,\:\:\: m =1, \ldots M,
         \end{equation}
         where ${\langle{w}_{m}\rangle}$, $m =1, \ldots M$ are
        expected values of $l$, the only information available about
        $l$ apart
        from the prior $r$. To apply minimum entropy principle to
        estimate posterior estimation $p$ of $l$, one has to determine
        the constraints for $p$ with respect to which we minimize
        $I(p\|r)$. Equivalently, by assuming that $p$ satisfies the
        constraints of the form (\ref{Equation:PT:InSS_PTrestated_2}),
        one has to determine the expected values ${\langle u_{m}
        \rangle}$, $m=1, \ldots, M$. 

         Now, as ${\langle u_{m} \rangle}, \:m=1,
         \ldots, M$ vary, one can show that $I_{q}(l\|p)$ has the minimum value when
         \begin{equation}
         \label{Equation:PTExpectedValueMatching_InClassicalCase}
         {\langle u_{m} \rangle} = {\langle w_{m} \rangle} \enspace,\:\:\: m = 1,
         \dots M.
         \end{equation}
        The proof is as
         follows~\cite{Shore:1981:PropertiesOfCrossEntropyMinimization}.
          Proceeding as in the 
         proof of Theorem~\ref{Theorm:PT:PTinClassicalCase_1}, we have
        {\setlength\arraycolsep{0pt}
        \begin{eqnarray}
        I(l\|p) &\:=\:&   I(l\|r) + \sum_{m=1}^{M} \beta_{m} \left\{ \int_{X} l(x) u_{m}(x)
         \, \ud \mu(x) \right\} + \ln \widehat{Z} \nonumber \\
                &\:=\:& I(l\|r) + \sum_{m=1}^{M} \beta_{m} {\langle w_{m} \rangle} 
                + \ln \widehat{Z} \qquad \qquad \mbox{(By (\ref{Equation:PT:InSS_PTrestated_4}))}
        \end{eqnarray}}
        Since the variation of $I(l\|p)$ with respect to ${\langle
         u_{m} \rangle}$ results in the variation of $I(l\|p)$ with
         respect to $\beta_{m}$ for any $m = 1, \ldots, M$, to find
         the minimum of $I(l\|p)$ one can solve
        \begin{displaymath}
        \frac{\partial}{\partial \beta_{m}} I_{q}(l\|p) =0 \enspace,
         \:\:\: m =1, \ldots M \enspace,
        \end{displaymath}
        which gives the solution as in
         (\ref{Equation:PTExpectedValueMatching_InClassicalCase}). 
    
         This property of expectation matching states that, for a
         distribution $p$ of the 
         form (\ref{Equation:PT:MinimumCrossEntropyDistribution}),
         $I(l\|p)$ is the smallest when the expected values of $p$ match
         those of $l$. In particular, $p$ is not only the
         distribution that minimizes $I(p\|r)$ but also minimizes
        $I(l\|p)$.

         We now restate the Theorem~\ref{Theorm:PT:PTinClassicalCase_1}
         which summarizes the above discussion. 
         \begin{theorem}
         \label{Theorm:PT:PTinClassicalCase_2}            
         Let $r$ be the prior distribution, and $p$ be the probability
         distribution that minimizes the relative-entropy subject to a
         set of constraints
         \begin{equation}
         \int_{X} u_{m}(x) p(x) \, \ud \mu(x) =
         \langle u_{m} \rangle \enspace, \qquad m = 
         1, \ldots , M.
         \end{equation}
         Let $l$ be any other distribution satisfying the constraints
         \begin{equation}
         \label{Equation:PT:In_PTinClassicalCase_2_ConstaintsFor_l}
         \int_{X} u_{m}(x) l(x) \, \ud \mu(x) =
         \langle w_{m} \rangle \enspace, \qquad m = 1, \ldots , M.
         \end{equation}
         Then
         \begin{enumerate} 
          \item $I_{1}(l\|p)$ is minimum only if (expectation matching property)
            \begin{equation}
            \label{Equation:PT:ConditionFor_TriangleEquality}
                {\langle u_{m} \rangle} = {\langle w_{m} \rangle}
         \enspace, \qquad m=1, \ldots M.
           \end{equation} 
          \item When (\ref{Equation:PT:ConditionFor_TriangleEquality})
         holds, we have
             \begin{equation}
             \label{Equation:PT:In_Theorem_PTinClassicalCase_2_TriangleEquality}        
                I(l\|r) = I(l\|p) + I(p\|r)
             \end{equation}     
         \end{enumerate}
         \end{theorem}

        By the above interpretation of triangle equality and analogy with
        the comparable situation in Euclidean 
        geometry, it is natural to call $p$, as defined by
        (\ref{Equation:PT:MinimumCrossEntropyDistribution}) as the
        projection of $r$ on the plane described by
        (\ref{Equation:PT:In_PTinClassicalCase_2_ConstaintsFor_l}).
        Csisz\'{a}r~\cite{Csiszar:1975:I-devergenceOfProbabilityDistributionsAndMinimizationProblems}
        has introduced a generalization of this notion to define the
        projection of $r$ on any convex set $\mathcal{E}$ of
        probability distributions. If $p \in \mathcal{E}$ satisfies
        the equation
        \begin{equation}
             I(p\|r) = \min_{s \in \mathcal{E}} I(s\|r) \enspace,
        \end{equation}
        then $p$ is called the projection of $r$ on
        $\mathcal{E}$.
        Csisz\'{a}r~\cite{Csiszar:1975:I-devergenceOfProbabilityDistributionsAndMinimizationProblems}
        develops a number of results about these projections
        for both finite and infinite dimensional spaces. In this
        paper, we will not consider this general approach.

  \subsection{The Case of $q$-Expectations}
        \noindent
         From the above discussion, it is clear that to derive the triangle
         equality of Tsallis relative-entropy minimization, one should
         first deduce the equivalent of expectation matching property
         in the nonextensive case.
         
         We state below and prove the Pythagoras theorem in
         nonextensive
         framework established by Dukkipati~et~al.~\cite{DukkipatiMurtyBhatnagar:2006:NonextensiveTriangleEquality}.   
         \begin{theorem}
         \label{Theorem:PT:PTinNonextensiveCase_q-Expectations}            
         Let $r$ be the prior distribution, and $p$ be the probability
         distribution that minimizes the Tsallis relative-entropy
         subject to a set of constraints
         \begin{equation}
         \label{Equation:PT:InNonExtPT1_ExpectationConstraints_For_p}
         \int_{X} u_{m}(x) p(x)^{q} \, \ud \mu(x) =
         {\langle u_{m} \rangle}_{q} \enspace, \qquad m = 
         1, \ldots , M.
         \end{equation}
         Let $l$ be any other distribution satisfying the constraints
         \begin{equation}
         \label{Equation:PT:InNonExtPT1_ExpectationConstraints_For_l}
         \int_{X} u_{m}(x) l(x)^{q} \, \ud \mu(x) =
         {\langle w_{m} \rangle}_{q} \enspace, \qquad m = 1, \ldots , M.
         \end{equation}
         Then
         \begin{enumerate}
          
          \item $I_{q}(l\|p)$ is minimum only if
              \begin{equation}
                \label{Equation:PT:ConditionFor_NonExtTri_q-Expect}
                  {\langle u_{m} \rangle}_{q} = \frac{ {\langle w_{m}
                   \rangle}_{q} }{ 1 - (1-q) I_{q}(l\|p) } \enspace,\:\:\: m =1,
                   \ldots M. 
              \end{equation}

          \item Under
               (\ref{Equation:PT:ConditionFor_NonExtTri_q-Expect}), we have
               \begin{equation}
               \label{Equation:PT:NonextensiveTriangleEquality_q-Expect}
                   I_{q}(l\|r) = I_{q}(l\|p) + I_{q}(p\|r) + (q-1)
               I_{q}(l\|p) I_{q}(p\|r) \enspace.
                \end{equation}
               
        \end{enumerate}
        \end{theorem}
        \proof{
        First we deduce the equivalent of expectation matching
        property in the nonextensive case. 
        That is, we would like to find the values of
        ${\langle{u}_{m}\rangle}_{q}$ for which $I_{q}(l\|p)$ is
        minimum. We write the following useful relations before we proceed
        to the derivation.

        We can write the generalized minimum entropy distribution
        (\ref{Equation:PT:GeneralizedMinimumCrossEntropyDistribution_2})
        as
         \begin{equation}
        \label{Equation:PT:GeneralizedMinimumCrossEntropyDistribution_3}   
         p(x) = \frac{\displaystyle e_{q}^{\ln_{q} r(x)} \otimes_{q}
        {e_{q}}^{- \sum_{m=1}^{M} \beta_{m} u_{m}(x)}}{\displaystyle
        \widehat{Z_{q}}}  
        = \frac{\displaystyle {e_{q}}^{- \sum_{m=1}^{M} \beta_{m}
        u_{m}(x) + \ln_{q} r(x) }}{\displaystyle \widehat{Z_{q}} }
        \enspace, 
         \end{equation}
        by using the relations $e_{q}^{\ln_{q} x } = x$ and $e_{q}^{x}
        \otimes_{q} e_{q}^{y} = e_{q}^{x+y}$.
        Further by using
        \begin{displaymath}
        \ln_{q}(xy) = \ln_{q} x + \ln_{q} y + (1-q) \ln_{q} x
        \ln_{q}y
        \end{displaymath}
        we can write
        (\ref{Equation:PT:GeneralizedMinimumCrossEntropyDistribution_3}) as
        \begin{equation}
        \label{Equation:PT:IntermediateRelationOfGeneralizedMinimumCrossEntropyDistribution}
        \ln_{q} p(x) + \ln_{q} \widehat{Z_{q}} +  (1-q)
        \ln_{q} p(x) \ln_{q} \widehat{Z_{q}} =  
        - \sum_{m=1}^{M} \beta_{m} u_{m}(x) + \ln_{q} r(x)
        \enspace. 
        \end{equation}
        By the property of
        $q$-logarithm
        \begin{equation}
        \label{Equation:PT:PropertyOflnq(x/y)}
        \ln_{q}\left(\frac{x}{y} \right)
              = y^{q-1}( \ln_{q}x - \ln_{q}y) \enspace,
        \end{equation}
        and by $q$-logarithmic representations of Tsallis entropy,
        \begin{displaymath}     
        S_{q} = - \int_{X} p(x)^{q} \ln_{q} p(x)\, \ud \mu(x) \enspace,
        \end{displaymath}
        one can verify that
        \begin{equation}
        \label{Equation:PT:RelationBetweenGeneralizedKLAndTsallis}
        I_{q}(p\|r) = - \int_{X}  {p(x)}^{q} \ln_{q}r(x) \, \ud \mu(x)- S_{q}(p) \enspace.
        \end{equation}

        With these relations in hand we proceed with the derivation. Consider
        \begin{displaymath}
        I_{q}(l\|p) = - \int_{X}  l(x) \ln_{q} \frac{p(x)}{l(x)} \, \ud \mu(x)  \enspace.
        \end{displaymath}
        By (\ref{Equation:PT:PropertyOflnq(x/y)}) we have
        {\setlength\arraycolsep{0pt}
        \begin{eqnarray}
        I_{q}(l\|p) &\: =\: & - \int_{X}  {l(x)}^{q} \Big[ \ln_{q} p(x) -
        \ln_{q} l(x) \Big] \, \ud \mu(x)  \nonumber \\
                   &\: =\: & I_{q}(l\|r) - \int_{X}  {l(x)}^{q}\Big[\ln_{q}
        p(x) - \ln_{q} r(x) \Big] \, \ud \mu(x) \enspace. 
        \end{eqnarray}}
        From
        (\ref{Equation:PT:IntermediateRelationOfGeneralizedMinimumCrossEntropyDistribution}),
        we get
        {\setlength\arraycolsep{0pt}
        \begin{eqnarray}
        I_{q}(l\|p) = I_{q}(l\|r)
        && + \int_{X}  {l(x)}^{q}  \left[ \sum_{m=1}^{M} \beta_{m} u_{m}(x)
        \right] \, \ud \mu(x)  \nonumber \\ 
         && + \ln_{q} \widehat{Z_{q}} \int_{X} {l(x)}^{q} \, \ud \mu(x)  \nonumber \\
        &&+ (1-q) \ln_{q}\widehat{Z_{q}} \int_{X} {l(x)}^{q} \ln_{q} p(x)
         \, \ud \mu(x)  \enspace.
        \end{eqnarray}}
        By using (\ref{Equation:PT:InNonExtPT1_ExpectationConstraints_For_l})
        and 
        (\ref{Equation:PT:RelationBetweenGeneralizedKLAndTsallis}), 
        {\setlength\arraycolsep{0pt}
        \begin{eqnarray}
        I_{q}(l\|p)  =  I_{q}(l\|r) + \sum_{m=1}^{M} \beta_{m}&&
        {\langle{{w}_{m}}\rangle}_{q}  + \ln_{q}
        \widehat{Z_{q}} \int_{X} {l(x)}^{q} \, \ud \mu(x) \nonumber \\ 
        && + (1-q) \ln_{q} \widehat{Z_{q}}
        \Big[ - I_{q}(l\|p) - S_{q}(l) \Big] \enspace,  
        \end{eqnarray}}
        and by the expression of Tsallis entropy $S_{q}(l) = \frac{1}{q-1} \left[ 1 - \int_{X}
        l(x)^{q}\, \ud \mu(x) \right] $, we have   
        \begin{equation}
        \label{Equation:MainIntermediateEqForNonextensiveTriangleEquality}
        I_{q}(l\|p) = I_{q}(l\|r)+ \sum_{m=1}^{M} \beta_{m}
        {\langle{{w}_{m}}\rangle}_{q} + \ln_{q} \widehat{Z_{q}}
        - (1-q) \ln_{q} \widehat{Z_{q}} I_{q}(l\|p) \enspace.
        \end{equation}
        Since the multipliers $\beta_{m},\:\ m=1,\ldots M $ are
        functions of the expected values ${\langle u_{m}
        \rangle}_{q}$, variations in the expected values are
        equivalent to variations in the multipliers. Hence, to find the
        minimum of $I_{q}(l\|p)$, we solve
        \begin{equation}
        \label{Equation:PT:InPTinNonextensiveCase_PartialDerivative_For_ExpectMatchingProperty}
        \frac{\partial}{\partial \beta_{m}} I_{q}(l\|p) =0 \enspace.
        \end{equation}

        By using thermodynamic equation 
        (\ref{Equation:PT:GeneralizedKL_ThermodyamicEquation_1}),
        solution of
        (\ref{Equation:PT:InPTinNonextensiveCase_PartialDerivative_For_ExpectMatchingProperty})  
        provides us with the expectation matching property in the
        nonextensive case as
        \begin{equation}
        \label{Equation:ClosedForm_Tsallis_ExpectedValueMatching}
        {\langle u_{m} \rangle}_{q} = \frac{ {\langle w_{m}
        \rangle}_{q} }{ 1 - (1-q) I_{q}(l\|p) } \enspace,\:\:\: m =1,
        \ldots M \enspace. 
        \end{equation}
        In the limit $q \rightarrow 1$ the above equation gives $
        {\langle u_{m} \rangle}_{1} = {\langle w_{m} \rangle}_{1}$
        which is the expectation matching property in the classical
        case. 

        Now, to derive the triangle equality for Tsallis
        relative-entropy minimization, we substitute the expression
        for ${\langle w_{m} \rangle}_{q}$,
        which is given by~(\ref{Equation:ClosedForm_Tsallis_ExpectedValueMatching}),
        in~(\ref{Equation:MainIntermediateEqForNonextensiveTriangleEquality}).
        And after some algebra one can arrive at
        (\ref{Equation:PT:NonextensiveTriangleEquality_q-Expect}).
        \endproof
        
        Note that the limit $q \rightarrow 1$
        in~(\ref{Equation:PT:NonextensiveTriangleEquality_q-Expect})
        gives the triangle equality in the classical 
        case~(\ref{Equation:PT:In_Theorem_PTinClassicalCase_2_TriangleEquality}).
        The two important cases which arise out
        of~(\ref{Equation:PT:NonextensiveTriangleEquality_q-Expect}) are, 
        \begin{eqnarray}
        I_{q}(l\|r) &\leq& I_{q}(l\|p) + I_{q}(p\|r) \:\:\:
        \mbox{when} \:  0 < q \leq 1 \enspace,\\
        I_{q}(l\|r) &\geq& I_{q}(l\|p) + I_{q}(p\|r) \:\:\:
        \mbox{when} \: 1 < q \enspace.
        \end{eqnarray}

        We refer to
        Theorem~\ref{Theorem:PT:PTinNonextensiveCase_q-Expectations} as
        nonextensive Pythagoras' theorem 
        and (\ref{Equation:PT:NonextensiveTriangleEquality_q-Expect}) as
        {\em nonextensive triangle equality}, whose pseudo-additivity property
        is consistent with the pseudo additivity of Tsallis
        relative-entropy
        {\setlength\arraycolsep{0pt}
        \begin{eqnarray}
        \label{Equation:KN:NonextensiveAdditivityOfTsallisRelativeEntropy}
        I_{q}(X_{1} \times Y_{1} \| X_{2} \times Y_{2} ) = 
        I_{q}(X_{1} &&\| X_{2}) + I_{q}(Y_{1} \| Y_{2}) \nonumber \\
        && + (q-1) I_{q}(X_{1} \| X_{2}) I_{q}(Y_{1} \| Y_{2}) \enspace,
        \end{eqnarray}}
        where $X_{1},X_{2}$ and $Y_{1},Y_{2}$ are r.vs such that $X_{1}$ and $Y_{1}$ are
        independent, and
        $X_{2}$ and $Y_{2}$ are independent respectively;
        hence is a natural generalization of triangle equality in
        the classical case.  

  \subsection{In the Case of Normalized $q$-Expectations}
         \noindent
         In the case of normalized $q$-expectation too, the Tsallis
        relative-entropy satisfies nonextensive triangle equality with
        modified conditions from the case of $q$-expectation
        values~\cite{DukkipatiMurtyBhatnagar:2006:NonextensiveTriangleEquality,Dukkipati:2006:PhdThesis}. 
         \begin{theorem}
         \label{Theorm:PT:PTinNonextensiveCase_Normalized-q-Expectations}            
         Let $r$ be the prior distribution, and $p$ be the probability
         distribution that minimizes the Tsallis relative-entropy
         subject to the set of constraints
        \begin{equation}
        \label{Equation:PT:Normalized-q-ExpectationConstraints_For-p}
        \frac{\int_{X} u_{m}(x) p(x)^{q} \, \ud \mu(x)}{\int_{X}
          p(x)^{q}\, \ud \mu(x)} = {\langle\langle u_{m}
         \rangle\rangle}_{q} \enspace, m =  
        1, \ldots , M.
        \end{equation}
        Let $l$ be any other distribution satisfying the constraints
        \begin{equation}
        \label{Equation:PT:Normalized-q-ExpectationConstraints_For-l}
        \frac{\int_{X} u_{m}(x) l(x)^{q} \, \ud \mu(x)}{\int_{X}
          l(x)^{q} \, \ud \mu(x)} =
        {\langle\langle w_{m} \rangle\rangle}_{q} \enspace, m = 1, \ldots , M.
        \end{equation}
        Then we have
               \begin{equation}
                \label{Equation:PT:NonExtTri_Normalized-q-Expect}
                   I_{q}(l\|r) = I_{q}(l\|p) + I_{q}(p\|r) + (q-1)
         I_{q}(l\|p) I_{q}(p\|r) \enspace,
                \end{equation}
        provided
        \begin{equation}
         \label{Equation:PT:ConditionFor_NonExtTri_Normalized-q-Expect} 
             {\langle\langle u_{m} \rangle\rangle}_{q} =
             {\langle\langle w_{m} \rangle\rangle}_{q}\, m=1, \ldots
             M. 
        \end{equation}  
        \end{theorem}
        
        \proof{
        From Tsallis minimum entropy distribution $p$ in the case
        of normalized $q$-expected values
        (\ref{Equation:PT:Tsallis_minimumRelativeEntropyDistribution_For_Normalizedq-Expectations_q-Expo}),
        we have
          {\setlength\arraycolsep{1pt}
          \begin{eqnarray}
            \label{Equation:PT:RelationBetween-p-and-r_InNormalized-q-expectationProof}
             \ln_{q} r(x) - \ln_{q} p(x) =  \ln_{q}
        \widehat{\overline{{Z}_{q}}} + &&(1-q) \ln_{q} p(x) 
            \ln_{q}\widehat{\overline{{Z}_{q}}} \nonumber \\
            && +  \frac{ \sum_{m=1}^{M} \beta_{m} \left( u_{m}(x) - {\langle\langle
              u_{m} \rangle\rangle}_{q} \right) }{\int_{X} p(x)^{q} \,
        \ud \mu(x)} \enspace.
          \end{eqnarray}}
          Proceeding as in the proof of
          Theorem~\ref{Theorem:PT:PTinNonextensiveCase_q-Expectations},
          we have
          \begin{equation}
            I_{q}(l\|p) = I_{q}(l\|r) - \int_{X} l(x)^{q}
            \Big[\ln_{q}p(x) - \ln_{q} r(x) \Big] \, \ud \mu(x) \enspace.
          \end{equation}
          From~(\ref{Equation:PT:RelationBetween-p-and-r_InNormalized-q-expectationProof}), 
          we obtain
        {\setlength\arraycolsep{1pt}
        \begin{eqnarray}
        I_{q}(l\|p) = &&I_{q}(l\|r) 
        + \ln_{q}\widehat{\overline{{Z}_{q}}} \int_{X}l(x)^{q} \,\ud
        \mu(x) \nonumber \\
        &&+ (1-q)  \ln_{q}\widehat{\overline{{Z}_{q}}}
               \int_{X} l(x)^{q} \ln_{q} p(x) \, \ud \mu(x) \nonumber \\
        &&+\frac{1}{\int_{X}p(x)^{q}\, \ud \mu(x)} \sum_{m=1}^{M}
               \beta_{m} 
               \int_{X} l(x)^{q} \left( u_{m}(x) - {\langle\langle
                 u_{m}\rangle\rangle}_{q} \right) \, \ud \mu(x)  \enspace.
               \nonumber\\
        \end{eqnarray}}
        By
        (\ref{Equation:PT:Normalized-q-ExpectationConstraints_For-l})
        the same can be written as
        {\setlength\arraycolsep{1pt}
        \begin{eqnarray}
         \label{Equation:PT:IntermediateEquationInTheorem_1_Tri-Norm-q-Expect}
        I_{q}(l\|p) = I_{q}(l\|r) 
        &&+ \ln_{q}\widehat{\overline{{Z}_{q}}} \int_{X}l(x)^{q} \,\ud
        \mu(x) \nonumber \\
        &&+ (1-q)  \ln_{q}\widehat{\overline{{Z}_{q}}}
               \int_{X} l(x)^{q} \ln_{q} p(x) \, \ud \mu(x) \nonumber \\
        &&+\frac{\int_{X}
               l(x)^{q} \, \ud \mu(x)}{\int_{X}p(x)^{q}\, \ud \mu(x)}   
                \sum_{m=1}^{M} \beta_{m}
               \left( {\langle\langle w_{m} \rangle\rangle}_{q} 
                - {\langle\langle u_{m} \rangle\rangle}_{q}
                \right) \enspace.
               \nonumber\\
        \end{eqnarray}}
        By using the relations
        \begin{displaymath}
          \int_{X} l(x)^{q} \ln_{q} p(x) \, \ud \mu(x) = - I_{q}(l\|p)
        - S_{q}(l) \enspace,
        \end{displaymath}
        and
        \begin{displaymath}
          \int_{X} l(x)^{q} \, \ud \mu(x) = (1-q) S_{q}(l) + 1 \enspace,
        \end{displaymath}
         (\ref{Equation:PT:IntermediateEquationInTheorem_1_Tri-Norm-q-Expect})
        can be written as
         {\setlength\arraycolsep{1pt}
        \begin{eqnarray}
        I_{q}(l\|p) = I_{q}(l\|r) && + 
        \ln_{q}\widehat{\overline{{Z}_{q}}} -  (1-q)
        \ln_{q}\widehat{\overline{{Z}_{q}}} I_{q}(l\|p) \nonumber\\ 
          &&+  
        \frac{\int_{X} l(x)^{q} \, \ud \mu(x)}{\int_{X}p(x)^{q}\, \ud \mu(x)}    
           \sum_{m=1}^{M} \beta_{m}
               \left( {\langle\langle w_{m} \rangle\rangle}_{q} 
                - {\langle\langle u_{m} \rangle\rangle}_{q}
                \right) \enspace.
        \end{eqnarray}}
         Finally using
        (\ref{Equation:PT:MinimumTsallisRelativeEntropy_Normalized-q-Expect})
        and 
         (\ref{Equation:PT:ConditionFor_NonExtTri_Normalized-q-Expect})
         we have the nonextensive triangle equality
        (\ref{Equation:PT:NonExtTri_Normalized-q-Expect}). 
         \endproof

        Note that in this case the minimum of $I_{q}(l\|p)$ is not
        guaranteed. Also the condition
        (\ref{Equation:PT:ConditionFor_NonExtTri_Normalized-q-Expect})
        for nonextensive triangle equality here is the same as the
        expectation value matching property in the classical case.


        
\section{Conclusions}
\label{Section:Conclusions}
        \noindent
        Phythagoras' theorem of relative-entropy plays an important
        role in geometrical approaches of statistical estimation
        theory like information geometry. In this paper we presented
        Pythagoras' theorem in the nonextensive case i.e., for Tsallis
        relative-entropy minimization. In our opinion, this result is yet another
        remarkable and consistent generalization shown by the Tsallis
        formalism.
        
        Now, equipped with the nonextensive Pythagoras' theorem in the
        generalized case of Tsallis, it is interesting to know the
        resultant geometry when we use generalized information
        measures and role of entropic index in the geometry.


\section*{References}

\bibliographystyle{unsrt}
\bibliography{papi}

\end{document}